\documentclass[conference]{IEEEtran}
\IEEEoverridecommandlockouts
\usepackage{cite}
\usepackage{amsmath,amssymb,amsfonts}
\usepackage{graphicx}
\usepackage{textcomp}
\usepackage{xcolor}
\usepackage{balance}
\usepackage{cite}
\usepackage{amsmath,amssymb,amsfonts}
\usepackage{amsmath}
\usepackage{amsthm}
\DeclareMathOperator*{\argmax}{argmax}

\usepackage{graphicx}
\usepackage{textcomp}
\usepackage{xcolor}
\usepackage{graphicx}
\usepackage{float}
\usepackage{subfigure}
\usepackage{amsmath}
\usepackage{amsfonts,amssymb}
\usepackage{mathrsfs}
\usepackage{mathtools}
\usepackage{bm}
\usepackage{multirow}
\usepackage{array}
\usepackage{amssymb}
\usepackage{amsmath}
\usepackage{cite}
\usepackage{url}
\usepackage{xcolor}
\usepackage{cite,graphicx,amsmath,amssymb,cases}
\usepackage{subfigure}
\usepackage{fancyhdr}
\usepackage{mdwmath}
\usepackage{mdwtab}
\usepackage{caption}
\usepackage{amsthm}
\usepackage{setspace}
\usepackage{bm}
\usepackage{mathtools}
\usepackage{dsfont}
\usepackage{bbm}
\usepackage{algorithm}
\usepackage{epstopdf}
\usepackage{algorithmic}

\newtheorem{theorem}{Theorem}

\newtheorem{lemma}{Lemma}

\newtheorem{corollary}{Corollary}

\makeatletter
\newcommand{\biggg}{\bBigg@{3}}
\newcommand{\Biggg}{\bBigg@{3.5}}
\makeatother
\def\BibTeX{{\rm B\kern-.05em{\sc i\kern-.025em b}\kern-.08em
    T\kern-.1667em\lower.7ex\hbox{E}\kern-.125emX}}
    \expandafter\def\expandafter\normalsize\expandafter{%
    \normalsize%
    \setlength\abovedisplayskip{4pt}%
    \setlength\belowdisplayskip{4pt}%
    \setlength\abovedisplayshortskip{2pt}%
    \setlength\belowdisplayshortskip{2pt}%
}
\begin{document}
\title{Exploiting Movable Antennas in Multicast Communications}
\author{\IEEEauthorblockN{Zhenqiao Cheng$^{\dag}$, Nanxi Li$^{\dag}$, Jianchi Zhu$^{\dag}$, Chongjun~Ouyang$^{\star}$, and Xingqi Zhang$^{\ddag}$}
$^\dag$6G Research Centre, China Telecom Beijing Research Institute, Beijing, 102209, China\\
$^{\star}$Queen Mary University of London, London, U.K. $^{\ddag}$University of Alberta, Edmonton, Canada\\
Email: $^{\dag}$\{chengzq, linanxi, zhujc\}@chinatelecom.cn, $^{\star}$c.ouyang@qmul.ac.uk, $^{\ddag}$xingqi.zhang@ualberta.ca}
\maketitle
\begin{abstract}
This article investigates the integration of movable antennas (MAs) into multicast communication systems. By discretizing the motion of the MAs, a novel MA-assisted multicast transmission architecture is formulated. An alternating optimization (AO) algorithm based on successive convex approximation is proposed to optimize the transmit beamforming and antenna positions. To gain further insights, the two-user case is examined, and a closed-form expression for the optimal beamformer is derived. On this basis, a low-complexity greedy search algorithm is developed to optimize the placement of the MAs. Furthermore, under the assumption of a line-of-sight propagation environment, a branch-and-bound algorithm is designed to determine the globally optimal antenna configuration with reduced complexity compared to exhaustive search. Numerical simulations confirm that the proposed methods effectively enhance the achievable multicast rate.
\end{abstract}
\section{Introduction}
\label{sec:intro}
Multiple-antenna technology has long been recognized as a cornerstone of modern wireless communication systems and continues to play a pivotal role in shaping the evolution of commercialized networks. In response to the increasingly stringent requirements of next-generation wireless networks, fluid antenna systems have emerged as a promising research direction \cite{Lu2025}. A fluid antenna refers to a versatile multiple-antenna system capable of dynamically altering its shape and position to reconfigure key parameters such as gain, radiation pattern, operating frequency, and more \cite{New2024}. This extreme flexibility distinguishes fluid antennas from conventional antenna systems and unlocks new potentials for performance optimization and transformative gains in wireless communications.

A notable implementation of fluid-antenna concept is the \emph{movable antenna (MA)} \cite{New2024,Zhu2023}. By connecting MAs to radio-frequency (RF) chains via flexible cables and employing real-time actuation through devices such as stepper motors or servos, these systems overcome the traditional positional constraints of fixed antennas \cite{Zhu2025}. This spatial flexibility enables MAs to reshape the wireless channel environment dynamically, thereby achieving enhanced transmission performance. For further insight, readers are referred to recent works in this domain \cite{New2024,Zhu2025}.

While existing studies have predominantly focused on the performance benefits of MAs in unicast transmission scenarios---where each user receives an independent data stream \cite{New2024,Zhu2025}---such approaches often encounter severe interference and elevated system complexity, particularly when the number of users scales. To address these limitations, multicast transmission, which enables simultaneous delivery of identical content to multiple users, has gained growing attention \cite{Golrezaei2013,Sidiropoulos2006,Jindal2006}. This paradigm is especially relevant in applications such as video conferencing, live broadcasting, and emergency alert systems, where disseminating high-volume, critical information efficiently is essential. Despite its practical relevance, the integration of MAs into multicast communications remains underexplored. Only a few recent works have considered this topic \cite{Gao2024,Kang2024}, and although they propose beamforming optimization strategies that demonstrate the promise of MAs, they fall short of providing comprehensive system-level design insights. In particular, the joint optimization of beamforming and antenna positioning has not been thoroughly addressed.

To bridge this gap, this paper investigates an MA-assisted multicast transmission framework within a multiple-input single-output (MISO) system. By modeling the positions of MAs as discrete variables constrained by mechanical motion steps, we formulate a novel architecture that adapts the antenna placement to enhance multicast rate performance. To efficiently solve the resulting joint design problem, we propose an alternating optimization (AO) algorithm based on successive convex approximation (SCA), which iteratively updates the transmit beamformer and the MA positions. To gain further insight, we analyze the two-user case and derive a closed-form optimal transmit beamformer. This analytical result then motivates a low-complexity greedy search method to optimize MA positioning. Additionally, for scenarios with line-of-sight (LoS) propagation, we develop a branch-and-bound (BAB) algorithm to compute the globally optimal MA configuration while significantly reducing computational burden compared to brute-force search. Numerical simulations validate the proposed framework, showing that MA-assisted multicast systems provide greater spatial degrees of freedom and superior multicast rate performance compared to conventional fixed-position antenna (FPA) systems.

\section{System Model}
\subsection{System Description}
\begin{figure}[!t]
 \centering
\setlength{\abovecaptionskip}{0pt}
\includegraphics[height=0.18\textwidth]{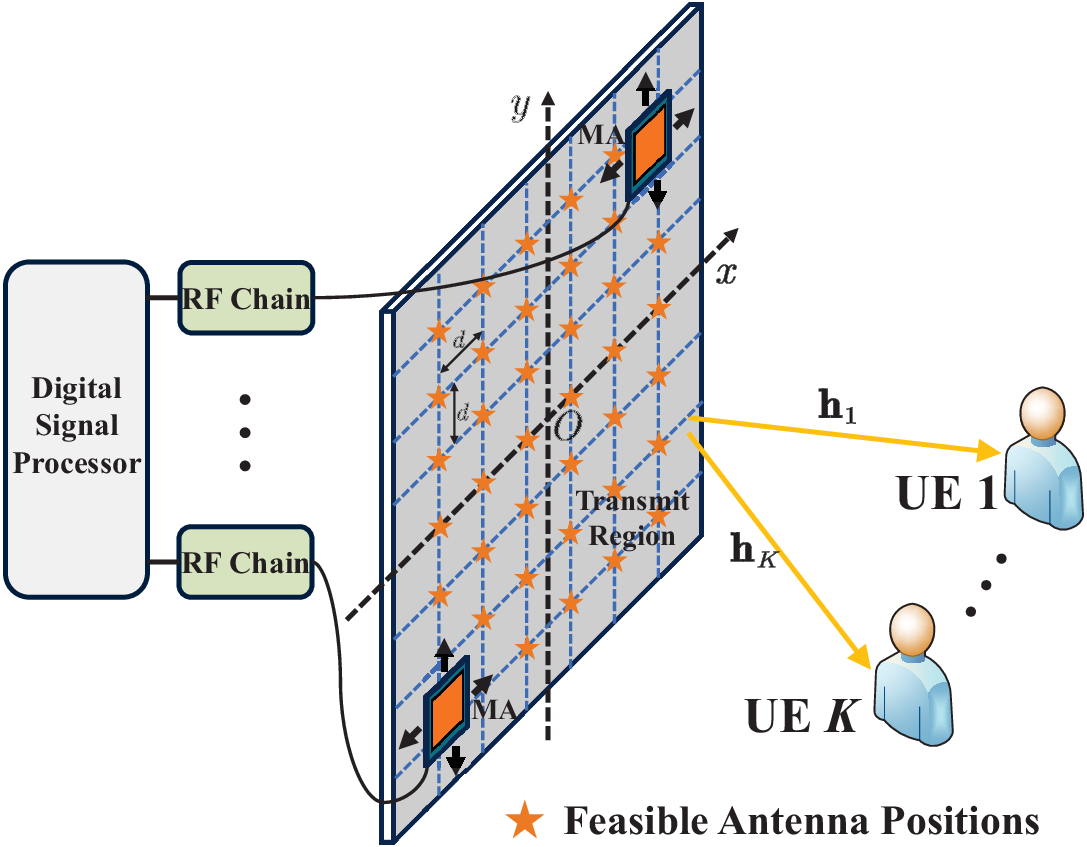}
\caption{Illustration of MA-aided multicast communications.}
\label{System_Model_Multicast}
\vspace{-10pt}
\end{figure}

We consider a multicast communication system, as illustrated in {\figurename} {\ref{System_Model_Multicast}}, where a transmitter equipped with $N$ MAs delivers a public message to $K$ single-antenna user equipments (UEs). The MAs are physically maneuvered via flexible cables within a predefined two-dimensional region ${\mathcal{C}}$ within the $x$-$y$ plane. In this architecture, each RF chain is connected to one MA. Due to limitations of practical electromechanical devices, MA displacements are restricted to horizontal or vertical movements with a fixed step size $d$, resulting in a quantized transmitter area. Let $M$ denote the total number of candidate discrete MA positions, which are defined by the set ${\mathcal{P}}\triangleq\{{\mathbf{p}}_m=[\tilde{x}_m,\tilde{y}_m]^{\mathsf{T}}\}_{m=1}^{M}$. These positions are spaced at exact intervals of $d$ along either axis, as shown in {\figurename} {\ref{System_Model_Multicast}}. The Cartesian coordinates of the $n$th transmit MA are denoted by ${\mathbf{t}}_n=[x_n, y_n]^{\mathsf{T}}$, and the feasible location set for each MA is given by $\mathcal{P}$, i.e., ${\mathbf{t}}_n\in{\mathcal{P}}$. For clarity, let ${\mathcal{T}}\triangleq\{{\mathbf{t}}_1 \ldots {\mathbf{t}}_N\}$ store the coordinates of the $N$ MAs.

The channel vector from the transmitter to UE $i$, which is denoted as ${\mathbf h}_{{k}}({\mathcal{T}})\in{\mathbbmss{C}}^{N\times1}$ for $k\in{\mathcal{K}}\triangleq\{1,\ldots,K\}$, depends on the signal propagation environment and the MA positions ${\mathcal{T}}$. Adopting the field-response channel model, we have ${\mathbf h}_{{k}}({\mathcal{T}})=[h_{k}({\mathbf{t}}_1)\ldots h_{k}({\mathbf{t}}_N)]^{\mathsf{T}}$ \cite{Ma2023}, where
\begin{align}\label{Channel_Model}
h_{k}(\mathbf{x})\triangleq\sum\nolimits_{\ell=1}^{L_k}\sigma_{\ell,k}{\rm{e}}^{{\rm{j}}\frac{2\pi}{\lambda}{\mathbf{x}}^{\mathsf{T}}{\bm\rho}_{\ell,k}},~k\in{\mathcal{K}}.
\end{align}
Here, $\lambda$ denotes the carrier wavelength, $\sigma_{\ell,k}$ is the complex gain of the $\ell$th path, and $L_k$ represents the number of resolvable paths. The vector ${\bm\rho}_{\ell,k}=[\sin{\theta_{\ell,k}}\cos{\phi_{\ell,k}},\cos{\theta_{\ell,k}}]^{\mathsf{T}}$ incorporates the elevation angle $\theta_{\ell,k}\in[0,\pi)$ and azimuth angle $\phi_{\ell,k}\in[0,\pi)$ of the $\ell$th path for the $k$th user. It is assumed that the transmitter can estimate the full channel state information (CSI) for each UE using existing methods \cite{Ma2023}. Furthermore, we consider quasi-static block-fading channels and analyze a specific fading block where the multipath components remain fixed across all locations in set $\mathcal{C}$.  

The transmitter precodes the public multicast signal ${\mathbf{s}}\in{\mathbbmss{C}}^{N\times1}$, derived from a normalized data symbol $x\in{\mathbbmss{C}}$ with $\mathbbmss{E}[|x|^2]=1$ and $\mathbbmss{E}[x^{\mathsf H}x]=0$, using a linear beamforming vector ${\mathbf{w}}\in{\mathbbmss{C}}^{N\times1}$. The received signal at each UE $k$ is then expressed as follows:
\begin{align}\label{System_model}
{y}_k={\mathbf{h}}_k^{\mathsf{H}}(\mathcal{T}){\mathbf{s}}+n_k={\mathbf{h}}_k^{\mathsf{H}}(\mathcal{T}){\mathbf{w}}x+n_k,~k\in{\mathcal{K}},
\end{align}
where $n_k\sim{\mathcal{CN}}(0,\sigma_k^2)$ represents the additive white Gaussian noise (AWGN) at UE $k$ with noise power $\sigma_k^2$. The multicast rate is given by \cite{Sidiropoulos2006,Jindal2006}
\begin{align}\label{Multicast_Capacity}
{\mathcal{R}}=\log_2(1+\min\nolimits_{k\in{\mathcal{K}}}\{\sigma_{k}^{-2}\lvert{\mathbf{h}}_{k}^{\mathsf{H}}(\mathcal{T}){\mathbf{w}}\rvert^2\}).
\end{align}
Unlike conventional FPAs, the MA-aided multicast rate in \eqref{Multicast_Capacity} depends on the MA positions $\mathcal{T}$. 
\subsection{Problem Formulation}
To characterize the rate limit of the MA-aided multicast system, we assume perfect channel state information (CSI) availability at both the transmitter and receivers. Our objective is to maximize the multicast rate ${\mathcal{R}}$ by jointly optimizing the MA positions $\mathcal{T}$ and the beamforming vector $\mathbf{w}$. The optimization problem is subject to discrete MA position constraints and a power budget, which is formulated as follows:
\begin{equation}\label{Rate_Maximization_Problem}\tag{${\mathcal{P}}_1$}
\begin{split}
\max_{\mathcal{T},{\mathbf{w}}}{\mathcal{R}}~
{\rm{s.t.}}&~\lVert\mathbf{w}\rVert^2\leq P,\\
&~{\mathbf{t}}_n\in{\mathcal{P}},n=1,\ldots,N,~{\mathbf{t}}_n\ne{\mathbf{t}}_{n'},n\ne n',
\end{split}
\end{equation}
where $P$ denotes the power budget. Problem \eqref{Rate_Maximization_Problem} is inherently non-convex due to the non-convex dependency of $\mathcal{R}$ on $({\mathcal{T}},{\mathbf{w}})$ and the discrete constraint ${\mathbf{t}}_n\in{\mathcal{P}}$. The tight coupling between $\mathbf{w}$ and $\mathcal{T}$ further complicates the optimization.
\section{A General Solution}
To decouple the interdependence between ${\mathbf{w}}$ and ${\mathcal{T}}$, we employ an AO framework. This method iteratively optimizes ${\mathbf{w}}$ and ${\mathcal{T}}$ while fixing the other variable.
\subsection{Transmit Beamforming Optimization}\label{A General Solution: Transmit Beamforming Optimization}
For fixed $\mathcal{T}$, problem \eqref{Rate_Maximization_Problem} reduces to a beamforming optimization over ${\mathbf{w}}$, which is formulated as follows: 
\begin{equation}\label{RaK_User_Transmit_Beamforming}
\begin{split}
\max_{{\mathbf{w}},r}~r~{\rm{s.t.}}&~{\mathbf{w}}^{\mathsf{H}}{\mathbf{h}}_{k}(\mathcal{T}){\mathbf{h}}_{k}^{\mathsf{H}}(\mathcal{T}){\mathbf{w}}\geq r\sigma_{k}^{2},~k\in\mathcal{K},\\
&~\lVert\mathbf{w}\rVert^2\leq P.
\end{split}
\end{equation}
Despite the non-convexity of \eqref{RaK_User_Transmit_Beamforming} due to the quadratic constraint ${\mathbf{w}}^{\mathsf{H}}{\mathbf{h}}_{k}(\mathcal{T}){\mathbf{h}}_{k}^{\mathsf{H}}(\mathcal{T}){\mathbf{w}}$, we resolve this via the successive convex approximation (SCA) technique. Specifically, we replace ${\mathbf{w}}^{\mathsf{H}}{\mathbf{h}}_{k}(\mathcal{T}){\mathbf{h}}_{k}^{\mathsf{H}}(\mathcal{T}){\mathbf{w}}$ with its first-order Taylor expansion-based lower bound. This yields the convex approximation as follows:
\begin{equation}\label{RaK_User_Transmit_Beamforming_SCA}
\begin{split}
\max\nolimits_{{\mathbf{w}},r}~r\quad{\rm{s.t.}}~&2\Re\{{\mathbf{w}}_{q}^{\mathsf{H}}{\mathbf{h}}_{k}(\mathcal{T}){\mathbf{h}}_{k}^{\mathsf{H}}(\mathcal{T}){\mathbf{w}}\}\\
&-\lvert{\mathbf{h}}_{k}^{\mathsf{H}}(\mathcal{T}){\mathbf{w}}_q\rvert^2
\geq r\sigma_{k}^{2},~k\in\mathcal{K},\\
&\lVert\mathbf{w}\rVert^2\leq P,
\end{split}
\end{equation}
where ${\mathbf{w}}_q$ represents the local point in the $q$th SCA iteration. Problem \eqref{RaK_User_Transmit_Beamforming_SCA} is a convex quadratically constrained program (QCP) solvable via standard convex solvers such as CVX. Algorithm \ref{alg0} outlines the SCA procedure, which guarantees monotonic improvement in the achievable rate and converges to a stationary-point solution of \eqref{RaK_User_Transmit_Beamforming}. The complexity scales as ${\mathcal{O}}(\ln\frac{1}{\varepsilon}K^{1.5}N^{4.5})$, where $\varepsilon$ is the convergence tolerance \cite{Chi2017}.

Notably, when $N\gg K$, say large-scale MU-MISO channel (e.g., in large-scale multiuser MISO channels), the SCA-based method incurs prohibitive computational complexity. To address this, we propose leveraging the signal subspace structure. Specifically, the optimal multicast beamformer must lie in the subspace spanned by the users' channel vectors $\{{\mathbf{h}}_{k}(\mathcal{T})\}_{k=1}^{K}$. This can be proven via contradiction. Assume that the optimal beamformer ${\mathbf{w}}^{\star}$ contains a component orthogonal to the subspace spanned by $\{{\mathbf{h}}_{k}(\mathcal{T})\}_{k=1}^{K}$. This orthogonal component consumes transmit power but contributes nothing to the received signal power at any user, violating the optimality of ${\mathbf{w}}^{\star}$. Thus, ${\mathbf{w}}^{\star}$ must reside in ${\rm{span}}(\{{\mathbf{h}}_{k}(\mathcal{T})\}_{k=1}^{K})$, allowing the parametrization:
\begin{align}\label{Optimal_Structure_Beamformer}
{\mathbf{w}}=\sum\nolimits_{k=1}^{K}\eta_k{\mathbf{h}}_{k}(\mathcal{T}),~\eta_k\in{\mathbbmss{C}}.
\end{align}
By optimizing the coefficients $\{\eta_k\}_{k=1}^{K}$ instead of $\mathbf{w}$ directly, the SCA complexity reduces to ${\mathcal{O}}(\ln\frac{1}{\varepsilon}K^{6})$, significantly lower than the original ${\mathcal{O}}(\ln\frac{1}{\varepsilon}K^{1.5}N^{4.5})$ when $N\gg K$. A detailed analysis of this subspace-aided optimization will be provided in an extended version of this work.

\begin{algorithm}[htbp]
\algsetup{linenosize=\tiny} \scriptsize
\caption{SCA-Based Approach for Solving \eqref{RaK_User_Transmit_Beamforming}}
\label{alg0}
\begin{algorithmic}[1]
\STATE initialize $\mathbf{w}_0$
    \REPEAT
    \STATE obtain ${\mathbf{{w}}}_{q+1}$ by solving problem \eqref{RaK_User_Transmit_Beamforming_SCA}
    \STATE update $q\leftarrow q+1$
    \UNTIL{the fractional decrease of the objective value of problem \eqref{RaK_User_Transmit_Beamforming} falls below a predefined threshold $\epsilon>0$}
\end{algorithmic}
\end{algorithm}

\subsection{Antenna Position Optimization}
For a fixed beamforming vector ${\mathbf{w}}$, the subproblem of \eqref{Rate_Maximization_Problem} for optimizing ${\mathcal{T}}$ is given as follows:
\begin{equation}\label{RaK_User_Antenna_Position}
\begin{split}
\max_{\mathcal{T}}&\min_{k\in{\mathcal{K}}}\{\sigma_{k}^{-2}\lvert{\mathbf{h}}_{k}^{\mathsf{H}}(\mathcal{T}){\mathbf{w}}\rvert^2\}\\
{\rm{s.t.}}~&{\mathbf{t}}_n\in{\mathcal{P}},n=1,\ldots,N,~{\mathbf{t}}_n\ne{\mathbf{t}}_{n'},n\ne n'.
\end{split}
\end{equation}
This problem is NP-hard due to the discrete constraints. To decouple dependencies among $\{{\mathbf{t}}_n\}_{n=1}^{N}$, we employ an element-wise AO framework. Specifically, each ${\mathbf{t}}_n$ is optimized sequentially while fixing all other antenna positions. The subproblem for ${\mathbf{t}}_n$ can be formulated as follows:
\begin{align}
\max\nolimits_{{\mathbf{t}}_n}&\gamma_{\rm{AO}}^{(n)}({\mathbf{t}}_n)\quad{\rm{s.t.}}~{\mathbf{t}}_n\in{\mathcal{P}},{\mathbf{t}}_n\ne{\mathbf{t}}_{n'},n\ne n',\label{MP_Pareto_Sub}
\end{align}
where $\gamma_{\rm{AO}}^{(n)}({\mathbf{x}})$ denotes the minimum signal-to-noise ratio (SNR) $\min_{k\in{\mathcal{K}}}\{\sigma_{k}^{-2}\lvert{\mathbf{h}}_{k}^{\mathsf{H}}(\mathcal{T}){\mathbf{w}}\rvert^2\}$, when ${\mathbf{t}}_n$ is assigned to $\mathbf{x}$ with all other positions fixed. This single-variable discrete optimization over the finite set $\mathcal{P}$ admits a globally optimal solution via exhaustive search with low complexity. The algorithm iteratively updates each ${\mathbf{t}}_n$ across all $N$ antennas until convergence. Algorithm \ref{Algorithm1} summarizes the full procedure. The computational complexity scales as ${\mathcal{O}}(I_{\rm{iter}} \sum_{k=1}^{K}L_k)$, where $I_{\rm{iter}}$ denotes the iteration count until convergence.

\begin{algorithm}[htbp]
\algsetup{linenosize=\tiny} \scriptsize
\caption{Element-wise Algorithm for Solving \eqref{RaK_User_Antenna_Position}}
\label{Algorithm1}
\begin{algorithmic}[1]
\STATE initialize the optimization variables
\REPEAT 
  \FOR{$n\in\{1,\ldots,N\}$}
      \STATE update ${\mathbf{t}}_n$ by solving problem \eqref{MP_Pareto_Sub} through a simple search
    \ENDFOR
\UNTIL{the fractional decrease of the objective value of problem \eqref{RaK_User_Antenna_Position} falls below a predefined threshold $\epsilon>0$}
\end{algorithmic}
\end{algorithm} 

\begin{algorithm}[htbp]
\algsetup{linenosize=\tiny} \scriptsize
\caption{Proposed AO-Based Algorithm for Problem \eqref{Rate_Maximization_Problem}}
\label{Algorithm3}
\begin{algorithmic}[1]
\STATE initialize $\{\mathbf{w}_0,{\mathcal{T}}_0\}$ and set $u=0$
    \REPEAT
    \STATE obtain ${\mathbf{{w}}}_{u+1}$ by solving problem \eqref{RaK_User_Transmit_Beamforming}
    \STATE obtain ${\mathcal{T}}_{u+1}$ by solving problem \eqref{RaK_User_Antenna_Position}
    \STATE update $u\leftarrow u+1$
    \UNTIL{the fractional decrease of the objective value of problem \eqref{Rate_Maximization_Problem} falls below a predefined threshold $\epsilon>0$}
\end{algorithmic}
\end{algorithm}

\subsection{Overall Algorithm: Convergence and Complexity Analysis}
Building on the preceding analysis, Algorithm \ref{Algorithm3} summarizes the proposed method for solving \eqref{Rate_Maximization_Problem}. Following \cite{Cheng2024}, Algorithm \ref{Algorithm3} generates a monotonically non-decreasing sequence of objective values for \eqref{Rate_Maximization_Problem}. Since the optimal value of \eqref{Rate_Maximization_Problem} is inherently upper-bounded by the transmit power and channel constraints, the algorithm is guaranteed to converge to a suboptimal solution. The computational complexity scales as ${\mathcal{O}}(I_{\rm{AO}}(\ln\frac{1}{\varepsilon}K^{1.5}N^{4.5}+I_{\rm{iter}} \sum_{k=1}^{K}L_k))$, where $I_{\rm{AO}}$ denotes the number of outer AO iterations required for convergence.

\section{Optimal Design: A Two-User Case Study}
The preceding section establishes a general framework for solving the joint beamforming and MA positioning problem \eqref{Rate_Maximization_Problem}. To derive deeper insights into optimal system design, we now analyze a simplified two-user case, i.e., $K=2$.
\subsection{Low-Complexity Two-Stage Design}
\subsubsection{Transmit Beamforming Optimization}
For a fixed MA position set $\mathcal{T}$, the optimal beamformer ${\mathbf{w}}^{\star}$ is derived from:
\begin{align}
{\mathbf{w}}^{\star}=\argmax_{\lVert\mathbf{w}\rVert^2\leq P}\min\{\lvert\sigma_1^{-1}{\mathbf{h}}_1^{\mathsf{H}}(\mathcal{T}){\mathbf{w}}\rvert,\lvert\sigma_2^{-1}{\mathbf{h}}_2^{\mathsf{H}}(\mathcal{T}){\mathbf{w}}\rvert\}.
\end{align}
The monotonic relationship between $\lvert{\mathbf{h}}_i^{\mathsf{H}}(\mathcal{T}){\mathbf{w}}\rvert^2$ ($i=1,2$) and $\lVert\mathbf{w}\rVert^2$ ensures that the optimal solution satisfies $\lVert\mathbf{w}^{\star}\rVert^2=P$. This permits the reformulation as follows:
\begin{subequations}\label{P1_Subproblem1_Trans}
\begin{align}
\min_{{\mathbf{p}},x}~&f({\mathbf{p}},x)=-x\\
{\rm{s.t.}}~&b_i({\mathbf{p}},x)=x-{\mathbf{p}}^{\mathsf{H}}\hat{\mathbf{h}}_i\hat{\mathbf{h}}_i^{\mathsf{H}}{\mathbf{p}}\leq0,i=1,2\\
&h({\mathbf{p}},x)=\lVert\mathbf{p}\rVert^2-1,
\end{align}
\end{subequations}
where $\hat{\mathbf{h}}_i\triangleq\frac{\sqrt{P}}{\sigma_i}\mathbf{h}_i(\mathcal{T})$ for $i=1,2$. Let $({\mathbf{p}}^{\star},x^{\star})$ denote the optimal solution. The corresponding beamformer is ${\mathbf{w}}^{\star}=\sqrt{P}{\mathbf{p}}^{\star}$, achieving a maximum multicast rate of $\log_2(1+x^{\star})$.

The optimal solution to problem \eqref{P1_Subproblem1_Trans} can be obtained from the Karush-Kuhn-Tucker (KKT) condition as follows \cite{Boyd2004}:
\begin{numcases}{}
  \nabla(-x)+\lambda\nabla(\lVert{\mathbf{p}}\rVert^2\!-\!1)+\mu_1\nabla b_{1}+\mu_2\nabla b_{2}={\mathbf{0}}, \label{KKT_1} \\
  \mu_1b_1=0, \mu_2b_2=0, \lambda\in{\mathbbmss{R}},\mu_1\geq0,\mu_2\geq0, \label{KKT_2}
\end{numcases}
where $\{\lambda,\mu_1,\mu_2\}$ are real-valued Lagrangian multipliers. From \eqref{KKT_1}, it can be shown that
\begin{numcases}{}
(\mu_1\hat{\mathbf{h}}_1\hat{\mathbf{h}}_1^{\mathsf{H}}+\mu_2\hat{\mathbf{h}}_2\hat{\mathbf{h}}_2^{\mathsf{H}}){\mathbf{p}}=\lambda{\mathbf{p}},\label{KKT_1_Dev1}\\
\mu_1+\mu_2=1\label{KKT_1_Dev2}.
\end{numcases}
Here, ${\mathbf{p}}$ is an eigenvector of $(\sum_{i=1}^{2}\mu_i\hat{\mathbf{h}}_i\hat{\mathbf{h}}_i^{\mathsf{H}})$ with eigenvalue $\lambda\geq0$. It follows from \eqref{KKT_1_Dev2} that $\mu_1$ and $\mu_2$ cannot be $0$ at the same time. Moreover, from \eqref{KKT_1_Dev1}, we have
\begin{equation}\label{KKT_2_Dev1}
\mu_1{\mathbf{p}}^{\mathsf{H}}\hat{\mathbf{h}}_1\hat{\mathbf{h}}_1^{\mathsf{H}}{\mathbf{p}}+
\mu_2{\mathbf{p}}^{\mathsf{H}}\hat{\mathbf{h}}_2\hat{\mathbf{h}}_2^{\mathsf{H}}{\mathbf{p}}=\lambda{\mathbf{p}}^{\mathsf{H}}{\mathbf{p}}=\lambda.
\end{equation}
The above results suggest that $\lambda>0$. 

Since $\mathbf{p}$ lies in ${\rm{span}}(\hat{\mathbf{h}}_1,\hat{\mathbf{h}}_2)$, we parameterize ${\mathbf{p}}=a\hat{\mathbf{h}}_1+b\hat{\mathbf{h}}_2$ \cite{Horn2012}. Particularly, we have ${\mathbf{p}}={\hat{\mathbf{h}}_1}/{\lVert\hat{\mathbf{h}}_1\rVert}$ for $\mu_2=0$ and $\mu_1>0$, and ${\mathbf{p}}={\hat{\mathbf{h}}_2}/{\lVert\hat{\mathbf{h}}_2\rVert}$ for $\mu_1=0$ and $\mu_2>0$. We then consider the case of $\mu_1>0$ and $\mu_2>0$, and it follows that
\begin{equation}\label{KKT_1_Dev3}
b_i=0\Leftrightarrow x={\mathbf{p}}^{\mathsf{H}}\hat{\mathbf{h}}_i\hat{\mathbf{h}}_i^{\mathsf{H}}{\mathbf{p}},~i=1,2.
\end{equation}
Substituting ${\mathbf{p}}=a\hat{\mathbf{h}}_1+b\hat{\mathbf{h}}_2$ into \eqref{KKT_1_Dev1} and \eqref{KKT_1_Dev3} gives
\begin{numcases}{}
a/b={{\mu_1}}/
{(\mu_2{\rm{e}}^{-{\rm{j}}\angle\alpha_{12}})},\label{KKT_1_Dev4}\\
\mu_1(\alpha_{1}+\alpha_{12}b/a)=\mu_2(\alpha_{12}^{*}a/b+\alpha_{2})=\lambda,\label{KKT_1_Dev5}
\end{numcases}
where $\alpha_{1}=\lVert\hat{\mathbf{h}}_{1}\rVert^2$, $\alpha_{2}=\lVert\hat{\mathbf{h}}_{2}\rVert^2$, and $\alpha_{12}=\hat{\mathbf{h}}_{1}^{\mathsf{H}}\hat{\mathbf{h}}_{2}$.

By combining \eqref{KKT_1_Dev2}, \eqref{KKT_1_Dev4}, and \eqref{KKT_1_Dev5}, we have
\begin{subequations}
\begin{align}
\mu_1=\frac{\alpha_2-\lvert\alpha_{12}\rvert}{\alpha_1+\alpha_2-2\lvert\alpha_{12}\rvert},~
\mu_2=\frac{\alpha_1-\lvert\alpha_{12}\rvert}{\alpha_1+\alpha_2-2\lvert\alpha_{12}\rvert},
\end{align}
\end{subequations}
which yields $\lambda=(\alpha_1\alpha_2-\lvert\alpha_{12}\rvert^2)(\alpha_1+\alpha_2-2\lvert\alpha_{12}\rvert)^{-1}$ and
\begin{equation}
{\mathbf{p}}=\frac{\mu_1\hat{\mathbf{h}}_1+\mu_2{\rm{e}}^{-{\rm{j}}\angle\alpha_{12}}\hat{\mathbf{h}}_2}
{\sqrt{\mu_1^2\alpha_1+\mu_2^2\alpha_2+2\mu_1\mu_2\lvert\alpha_{12}\rvert}}.
\end{equation}
Notice that $\lambda=\mu_1^2\alpha_1+\mu_2^2\alpha_2+2\mu_1\mu_2\lvert\alpha_{12}\rvert$, and it follows that ${\mathbf{p}}={\mu_1}/{\sqrt{\lambda}}\hat{\mathbf{h}}_1+{\mu_2}/{\sqrt{\lambda}}{\rm{e}}^{-{\rm{j}}\angle\alpha_{12}}\hat{\mathbf{h}}_2$. Taken together, the optimal beamforming vector for a given $\mathcal{T}$ is given as follows:
\begin{equation}\label{Optimal_Beamformer_Solution}
{\mathbf{w}}^{\star}=\left\{
\begin{array}{ll}
{\sqrt{P}{{\mathbf{h}}_i(\mathcal{T})}}/{\lVert{\mathbf{h}}_1(\mathcal{T})\rVert}             & {\alpha_i\leq\lvert\alpha_{12}\rvert}\\
\frac{\sqrt{P}\mu_1{\mathbf{h}}_{1}(\mathcal{T})}{\sigma_1\sqrt{\lambda}\lVert{\mathbf{h}}_1(\mathcal{T})\rVert}+
\frac{\sqrt{P}\mu_2{\mathbf{h}}_{2}(\mathcal{T}){\rm{e}}^{-{\rm{j}}\angle\alpha_{12}}}{\sigma_2\sqrt{\lambda}\lVert{\mathbf{h}}_2(\mathcal{T})\rVert}           & {\rm{Else}}
\end{array} \right.,
\end{equation}
with $i\in\{1,2\}$. The optimal $\mathbf{w}^{\star}$ resides in ${\rm{span}}({\mathbf{h}}_{1}(\mathcal{T}),{\mathbf{h}}_{2}(\mathcal{T}))$, consistent with the subspace property discussed in Section \ref{A General Solution: Transmit Beamforming Optimization}. 
\subsubsection{Antenna Position Optimization}
The preceding analysis implies that the maximum multicast rate is given by
\begin{equation}\label{Optimal_Multicast_Rate}
{\mathcal{R}}_1(\mathcal{T})=\left\{
\begin{array}{ll}
\log_2(1\!+\!\alpha_i)           & {\alpha_i\leq\lvert\alpha_{12}\rvert}\\
\log_2(1\!+\!\frac{\alpha_1\alpha_2-\lvert\alpha_{12}\rvert^2}{\alpha_1+\alpha_2-2\lvert\alpha_{12}\rvert})             & {\rm{Else}}
\end{array} \right.\!\!\!,
\end{equation}
with $i\in\{1,2\}$, which features a function of $\mathcal{T}$. This reformulation transforms problem \eqref{Rate_Maximization_Problem} into the equivalent optimization:
\begin{equation}\label{P1_Transform}
\begin{split}
\max_{\mathcal{T}}~{\mathcal{R}}_1(\mathcal{T})~{\rm{s.t.}}~{\mathbf{t}}_n\in{\mathcal{P}},{\mathbf{t}}_n\ne{\mathbf{t}}_{n'},n\ne n'.
\end{split}
\end{equation}
While problem \eqref{P1_Transform} is more tractable than \eqref{Rate_Maximization_Problem}, it retains non-convexity and NP-hardness due to discrete constraints. An exhaustive search for the optimal solution remains computationally prohibitive for large-scale systems.

To tackle this complexity, we propose a greedy search algorithm that iteratively approximates the optimal MA configuration. The algorithm operates over $N$ steps. At the $n$th step, the candidate antenna position $J_n$ is selected from the set $\mathcal J$ to maximize the incremental gain in ${\mathcal{R}}_1(\cdot)$ \cite{Gershman2004}:
\begin{align}\label{Greedy_Update}
J_n=\argmax\nolimits_{j\in{\mathcal J}}({\mathcal{R}}_1({\mathcal{S}}_{n-1}\cup\{{\mathbf{p}}_j\})-{\mathcal{R}}_1({\mathcal{S}}_{n-1})),
\end{align}
where ${\mathcal{S}}_{n-1}$ denotes the set of selected antenna coordinates after $n-1$ iterations. This greedy selection process is summarized in Algorithm \ref{alg4}, which reduces computational complexity from exponential to polynomial scaling. 

\begin{algorithm}[htbp]
\algsetup{linenosize=\tiny} \scriptsize
\caption{Greedy search-based approach for solving \eqref{Rate_Maximization_Problem}}
\label{alg4}
\begin{algorithmic}[1]
\STATE ${\mathcal J}=\{1,\ldots,M\}$, $n=0$, ${\mathcal{S}}_n=\emptyset$
    \REPEAT
    \STATE Set $n=n+1$ and update $J_n$ based on \eqref{Greedy_Update}
    \STATE Update ${\mathcal{S}}_n$ as ${\mathcal{S}}_{n-1}\cup J_n$
    \UNTIL{$n=N$}
\STATE Set the MA position set as ${\mathcal{T}}=\{{\mathbf{p}}_i\}_{i\in{\mathcal{S}}_N}$
\STATE Utilize \eqref{Optimal_Beamformer_Solution} to design the beamforming vector $\mathbf w$ for the given MA positions ${\mathcal{T}}$
\end{algorithmic}
\end{algorithm}

Note that ${\mathcal{R}}_1(\cdot)$ in \eqref{Optimal_Multicast_Rate} is a piecewise function. For practical implementation of Algorithm \ref{alg4}, the objective function can be evaluated as ${\mathcal{R}}_1(\cdot)$ as $\log_2(1+\alpha_i)$ and $\log_2(1+\frac{\alpha_1\alpha_2-\lvert\alpha_{12}\rvert^2}{\alpha_1+\alpha_2-2\lvert\alpha_{12}\rvert})$, respectively, and return the feasible solution of $\mathcal{T}$ that achieves the maximum objective value. The proposed \emph{two-stage algorithm} (stage 1: optimizing $\mathcal{T}$; stage 2: calculating $\mathbf{w}^{\star}$) \emph{eliminates AO loops}. This design reduces complexity significantly compared to the general framework in Algorithm \ref{alg4}.

\subsection{Special Case: LoS Transmission}
We analyze LoS propagation as a special case of problem \eqref{Rate_Maximization_Problem} with $L_1=L_2=1$. Assuming equal path loss for both users ( $\lvert\sigma_{1,1}\rvert^2=\lvert\sigma_{1,2}\rvert^2=\kappa$), which occurs when the UEs are equidistant from the transmitter, and equal noise power ($\sigma_1^2=\sigma_2^2=\sigma^2$), the multicast rate in \eqref{Optimal_Multicast_Rate} simplifies to
\begin{equation}\label{Optimal_Multicast_Rate_LOS}
{\mathcal{R}}_1(\mathcal{T})\!=\!\left\{
\begin{array}{ll}
\!\!\!\log_2(1\!+\!\frac{pN\kappa}{\sigma^2})           & {\alpha_1~{\text{or}}~\alpha_2\!\leq\!\lvert\alpha_{12}\rvert}\\
\!\!\!\log_2(1\!+\!\frac{p\kappa(N+\lvert{\mathbf{a}}_1^{\mathsf{H}}(\mathcal{T}){\mathbf{a}}_2(\mathcal{T})\rvert)}{2\sigma^2})             & {\rm{Else}}
\end{array} \right.\!\!\!,
\end{equation}
where ${\mathbf{a}}_i(\mathcal{T})\triangleq[{\rm{e}}^{{\rm{j}}\frac{2\pi}{\lambda}{\mathbf{t}}_1^{\mathsf{T}}{\bm\rho}_{1,i}}\ldots {\rm{e}}^{{\rm{j}}\frac{2\pi}{\lambda}{\mathbf{t}}_N^{\mathsf{T}}{\bm\rho}_{1,i}}]^{\mathsf{T}}$ for $i\in\{1,2\}$.

Define ${\mathbf{a}}\triangleq[a_1,\ldots,a_M]^{\mathsf{T}}\in\{0,1\}^{M\times1}$, where $a_m=1$ indicates an MA is placed at candidate position ${\mathbf{p}}_m$, and $a_m=0$ otherwise. This yields ${\mathcal{T}}=\{{\mathbf{p}}_m|a_m=1\}$ and
\begin{align}
\lvert{\mathbf{a}}_1^{\mathsf{H}}(\mathcal{T}){\mathbf{a}}_2(\mathcal{T})\rvert^2={{\mathbf a}^{\mathsf T}{\mathbf Q}{\mathbf a}},
\end{align}
where ${\mathbf Q}=\Re\{{\mathbf g}\}\Re\{{\mathbf g}^{\mathsf T}\}+\Im\{{\mathbf g}\}\Im\{{\mathbf g}^{\mathsf T}\}$, ${\mathbf g}={\mathbf g}_{1}^{*}\odot{\mathbf g}_{2}$, and ${\mathbf{g}}_i=[{\rm{e}}^{{\rm{j}}\frac{2\pi}{\lambda}{\mathbf{p}}_1^{\mathsf{T}}{\bm\rho}_{1,i}}\ldots {\rm{e}}^{{\rm{j}}\frac{2\pi}{\lambda}{\mathbf{p}}_M^{\mathsf{T}}{\bm\rho}_{1,i}}]^{\mathsf{T}}$ for $i=1,2$. Under these simplifications, problem \eqref{P1_Transform} reduces to the following:
\begin{align}\label{P2_BAB_Problem}
{\mathbf a}^{\star}=\argmax\nolimits_{a_m\in\{0,1\},\sum_{m=1}^{M}a_m=N}{{\mathbf a}^{\mathsf T}{\mathbf Q}{\mathbf a}}.
\end{align}
To solve \eqref{P2_BAB_Problem} efficiently, we propose a BAB algorithm that systematically \emph{prunes} suboptimal solutions, achieving lower complexity than exhaustive search.

Let $\mathcal S$ and $\mathcal{J}$ denote the index sets of selected MA positions and candidate positions, respectively. Initially, $\mathcal S$ is empty, and ${\mathcal J}={\mathcal J}_0\triangleq\{1,\ldots,M\}$. Let ${\mathbf a}_n$ represent the binary position determination vector after selecting $n$ MAs, with $f_n={{\mathbf a}_n^{\mathsf T}{\mathbf Q}{\mathbf a}}_n$ as the corresponding objective value. When selecting the $(n+1)$th position index $J_{n+1}$, the sets update as ${\mathcal S}\leftarrow{\mathcal S}\cup\{J_{n+1}\}$ and ${\mathcal J}\leftarrow{\mathcal J}\setminus\{J_{n+1}\}$. The objective $f_{n+1}$ updates recursively:
\begin{align}
f_{n+1}=f_{n}+2[{{\mathbf Q}{\mathbf a}_n^{\mathsf T}}]_{J_{n+1}}+[{\mathbf Q}]_{J_{n+1},J_{n+1}},
\end{align}
where $f_0=0$ and ${\mathbf a}_0$ is the zero vector. Here, $[]_i$ and $[]_{i,j}$ denote the $i$th element of a vector and the $(i,j)$th element of a matrix, respectively.

To align with the standard BAB framework \cite{Narendra1977}, which excels at optimizing monotonic functions, we reformulate the objective as follows:
\begin{align}
\overline{f}_{n}=f_{n}-2n{\mathsf X}-n{\mathsf Y},
\end{align}
where ${\mathsf X}=\max\nolimits_{j\in{\mathcal J}_0}[{\ddot{\mathbf Q}{\mathbf 1}^{\mathsf T}}]_{j}$ and ${\mathsf Y}=\max\nolimits_{j\in{\mathcal J}_0}[{\mathbf Q}]_{j,j}$, with $\ddot{\mathbf Q}$ denoting the element-wise absolute value of ${\mathbf Q}$. The modified objective updates recursively as follows:
\begin{equation}
\overline{f}_{n+1}=\overline{f}_{n}\!+\!2[{{\mathbf Q}{\mathbf a}_n^{\mathsf T}}]_{J_{n+1}}\!-\!2{\mathsf X}
\!+\![{\mathbf Q}]_{J_{n+1},J_{n+1}}\!-\!{\mathsf Y}.
\end{equation}
Since $[{{\mathbf Q}{\mathbf a}_n^{\mathsf T}}]_{J_{n+1}}<{\mathsf X}$ and $[{\mathbf Q}]_{J_{n+1},J_{n+1}}\leq{\mathsf Y}$, it holds that $\overline{f}_{n+1}<\overline{f}_{n}$, i.e., $\overline{f}_{n}$ decreases monotonically with $n$. Recall that the final objective $\overline{f}_{N}$ relates to ${f}_{N}$ through a constant offset $2L{\mathsf X}+N{\mathsf Y}$, which yields
\begin{equation}
\begin{split}
{\mathbf a}^{\star}=\argmax\nolimits_{\mathbf{a}}{f_{N}}=\argmax\nolimits_{\mathbf{a}}{\overline{f}_{N}}.
\end{split}
\end{equation}

\begin{figure}[!t]
 \centering
\setlength{\abovecaptionskip}{0pt}
\includegraphics[height=0.1\textwidth]{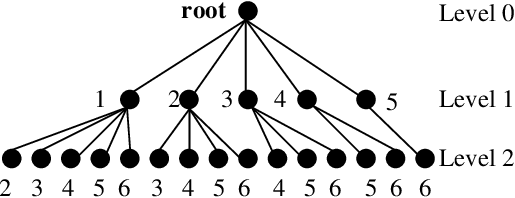}
\caption{A case of search tree for $M=6$ and $N=2$.}
\label{Tree_Model}
\end{figure}

The BAB process can be visualized as a tree search. {\figurename} {\ref{Tree_Model}} exhibits a case for $M=6$ and $N=2$, where each tree node represents a candidate position index. A root-to-leaf path corresponds to a selected MA subset, and the goal is to identify the path maximizing $\overline{f}_{N}$. By pruning suboptimal branches early, BAB reduces the search space significantly compared to exhaustive enumeration.

We initialize a global lower bound $B=-\infty$ and employ a best-first search strategy. The algorithm starts at the root node and iteratively explores subsequent levels of the search tree. For each node at level $n$, the corresponding objective value $\overline{f}_{n}$ is evaluated. If $\overline{f}_{n}<B$, the monotonically decreasing nature of $\overline{f}_{n}$ ensures that all child nodes of the current node will yield $\overline{f}_{n+1}<B$. Consequently, these child nodes are pruned to avoid redundant computations. If $\overline{f}_{n}>B$, the child nodes are prioritized and explored in ascending order of their predicted $\overline{f}_{n+1}$ values. After evaluating or pruning all nodes in the current subtree, the algorithm backtracks to parent nodes to explore alternate branches. This process continues until all nodes in the search tree are either examined or pruned, guaranteeing global optimality.

\begin{algorithm}[htbp]
\algsetup{linenosize=\tiny} \scriptsize
\caption{Optimal BAB-based approach for solving \eqref{P2_BAB_Problem}}
\label{alg2}
\begin{algorithmic}[1]
\STATE $B=-\infty$, $n=0$, $\overline{{f}}_{{{n}}}=0$, ${\mathbf u}={\mathbf 0}$, $K=0$
\IF{$n=L-1$}
\STATE ${c_k}=\overline{{f}}_{{{n}}}+2[{{\mathbf Q}{\mathbf u}^{\mathsf T}}]_{k}-2{\mathcal X}+[{\mathbf Q}]_{k,k}-{\mathcal Y}, \forall{k}\in{\mathcal{I}}_{n,{K}}$
\IF{$\max_{{k}\in{\mathcal{I}}_{n,{K}}}{c_k}>B$}
\STATE Set $[{\mathbf{u}}]_{\argmax_{{k}\in{\mathcal{I}}_{n,{K}}}{c_k}}=1$
\STATE Update the global lower bound $B=\max_{{k}\in{\mathcal{I}}_{n,{K}}}{c_k}$
\ENDIF
\ELSE
\STATE ${c_k}=\overline{{f}}_{{{n}}}+2[{{\mathbf Q}{\mathbf u}^{\mathsf T}}]_{k}-2{\mathcal X}+[{\mathbf Q}]_{k,k}-{\mathcal Y}, \forall{k}\in{\mathcal{I}}_{n,{K}}$
\STATE Sort ${c_k}, \forall{k}\in{\mathcal{I}}_{n,{K}}$ in a descending order to get an ordered index vector ${\mathbf{k}}$
\FOR{$i=1:\left|{\mathcal{I}}_{n,{K}}\right|$}
\IF{$c_{[{\mathbf{k}}]_i}>B$}
\STATE Set $[{\mathbf{v}}]_{[{\mathbf{k}}]_i}=1$, $\overline{{f}}_{{{n+1}}}=c_{[{\mathbf{k}}]_i}$, and $K=[{\mathbf{k}}]_i$
\STATE Set $n={n+1}$, and go to line 2
\ELSE
\STATE Break the loop
\ENDIF
\ENDFOR
\ENDIF
\STATE Set the position determination vector as ${\mathbf a}={\mathbf u}$
\STATE Utilize \eqref{Optimal_Beamformer_Solution} to design the beamforming vector $\mathbf w$ for the given MA positions ${\mathcal{T}}$
\end{algorithmic}
\end{algorithm}

Compared to exhaustive search, the BAB algorithm reduces computational complexity significantly through systematic pruning. The pseudocode for solving problem \eqref{P2_BAB_Problem} is summarized in Algorithm \ref{alg2}, where ${\mathcal{I}}_{l,{K}}$ denotes the child node indices of node $K$ at level $l$, with ${\mathcal{I}}_{0,{0}}=\{1,\ldots,M-N+1\}$. The computational complexity scales as ${\mathcal O}(N_{\rm{v}}N)$, where $N_{\rm{v}}$ represents the number of visited nodes during the search.
\section{Numerical Results}
We validate the efficacy of the proposed algorithms through simulations. Unless stated otherwise, we set $M=25$, $K=5$, $N=4$, $P=10$ dBm, and $\sigma_k^2=-95$ dBm ($\forall k\in{\mathcal{K}}$). The transmit area is discretized into a $\sqrt{M}\times\sqrt{M}$ grid with uniform spacing $d = \frac{\lambda}{2}$, as shown in {\figurename} {\ref{System_Model_Multicast}}. Users are uniformly distributed within a hexagonal cell of radius $150$ m. The free-space path loss for UE $k$ follows $-10\log_{10}{\mu_k}=92.5 + 20 \log_{10}[f_0(\text{GHz})] + 20 \log_{10}[d_k({\text{km}})]$, where $f_0 = 5$ GHz is the carrier frequency and $d_k$ denotes the BS-to-UE $k$ distance. Each channel realization assumes $L_k=4$ resolvable paths with complex gains $\sigma_{\ell,k}\sim{\mathcal{CN}}(0,\frac{\mu_k}{L_k})$ ($\forall \ell,k$). Elevation and azimuth angles are uniformly randomized within $[0, \pi]$. For benchmarking, we compare against a FPA system with $N$ antennas arranged in a uniform linear array spaced at $\frac{\lambda}{2}$, centered at the origin and aligned with the $x$-axis. All results are averaged over $100$ independent channel realizations with randomized initial optimization variables.

\begin{figure}[!t]
 \centering
\setlength{\abovecaptionskip}{0pt}
\includegraphics[width=0.35\textwidth]{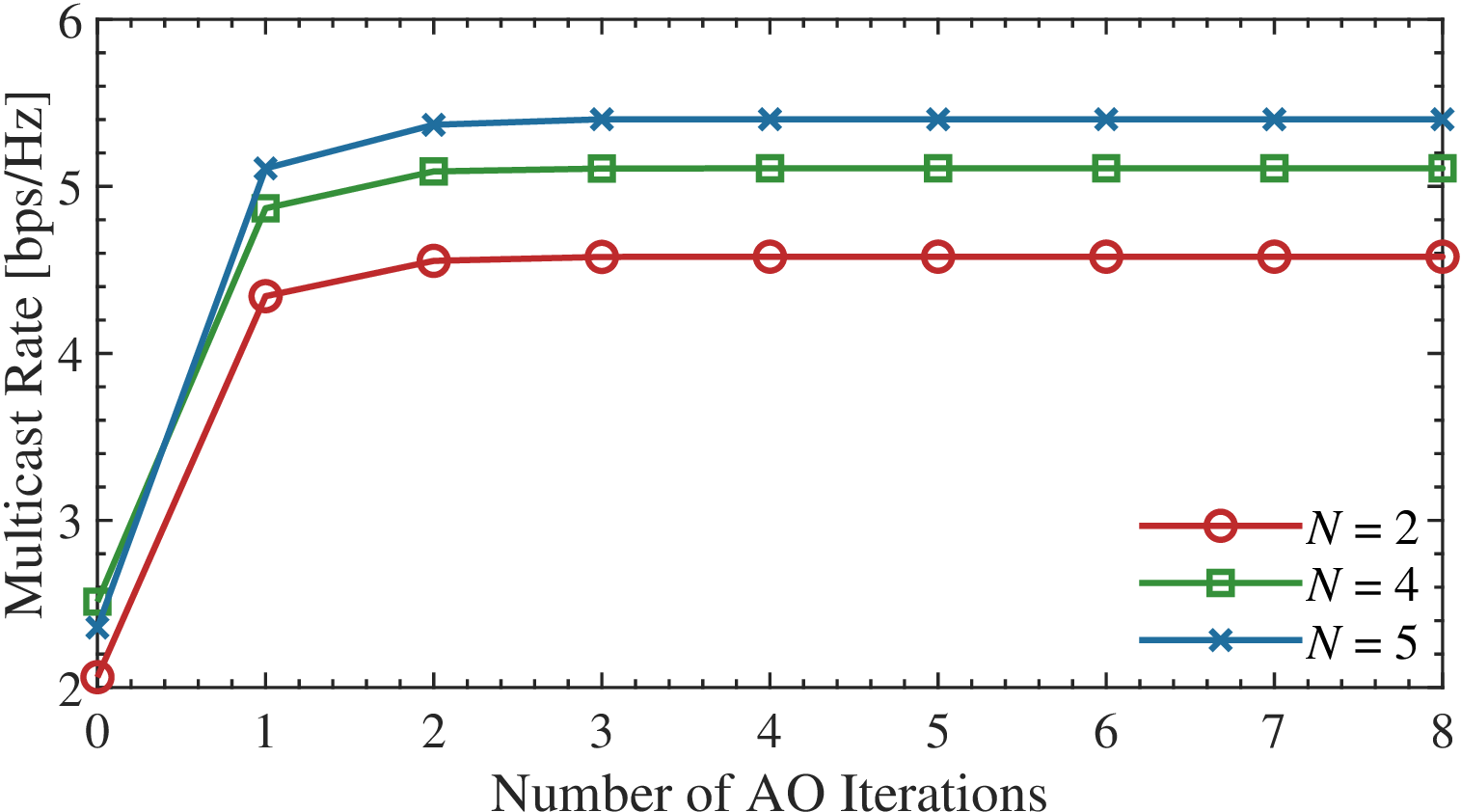}
\caption{Multicast rate vs. the number of AO iterations.}
\label{fig_AO_Complexity}
\end{figure}

\subsection{General Case}\label{Section:Numerical Results:General Case}
{\figurename} {\ref{fig_AO_Complexity}} demonstrates the convergence behavior of the proposed AO-based method for solving \eqref{Rate_Maximization_Problem}. The multicast rate exhibits rapid growth with the number of iterations, achieving convergence within approximately $5$ iterations. This highlights the computational efficiency of the proposed AO framework. Furthermore, increasing the number of MAs, i.e., $N$, enhances the multicast rate, as anticipated, due to the higher array gain from additional antennas.

\begin{figure}[!t]
 \centering
\setlength{\abovecaptionskip}{0pt}
\includegraphics[width=0.35\textwidth]{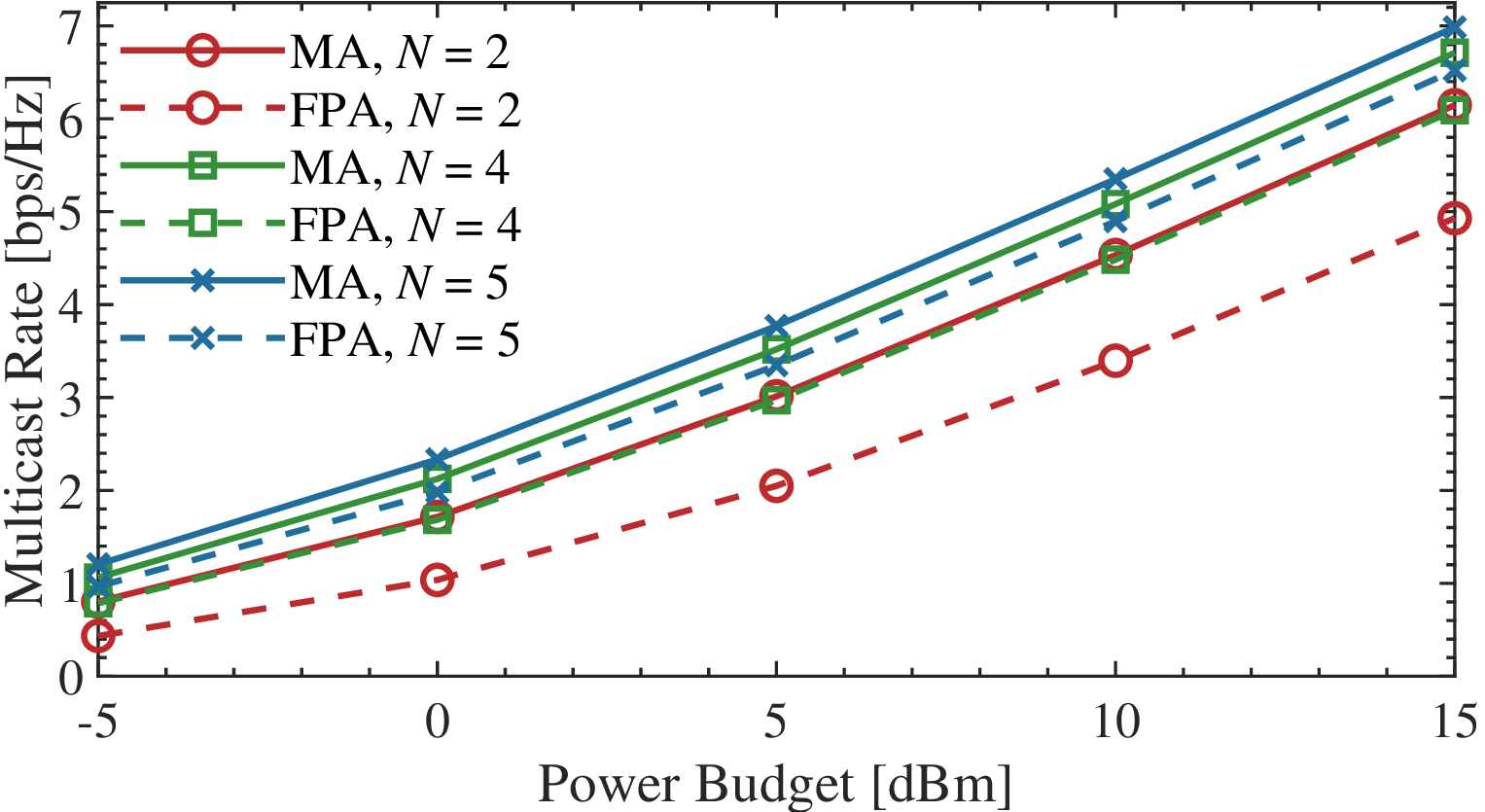}
\caption{Multicast rate vs. the power budget. $K=5$}
\label{fig_Rate_Power}
\end{figure}

\begin{figure}[!t]
 \centering
\setlength{\abovecaptionskip}{0pt}
\includegraphics[width=0.35\textwidth]{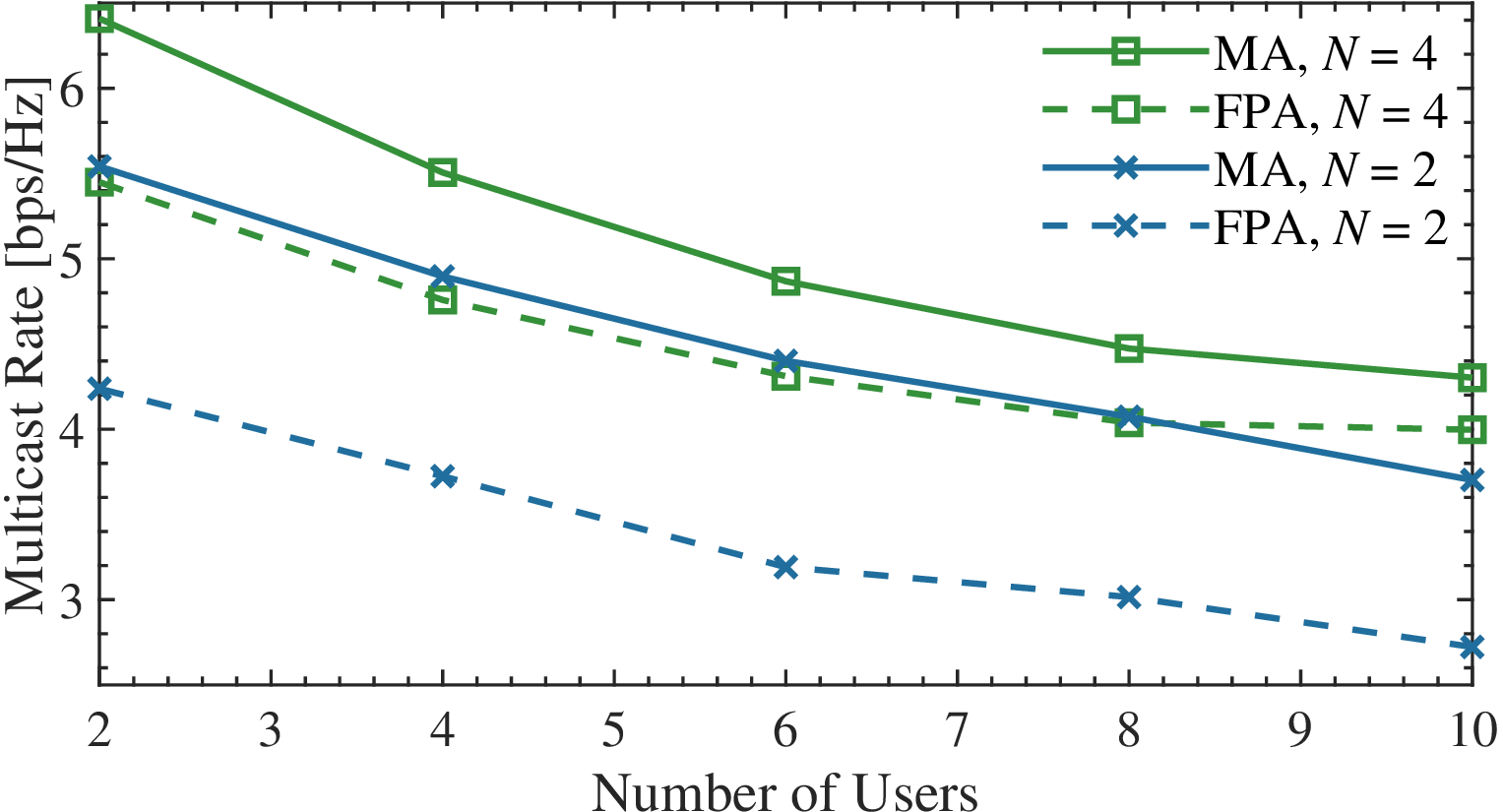}
\caption{Multicast rate vs. the number of users.}
\label{fig_Rate_User}
\end{figure}

{\figurename} {\ref{fig_Rate_Power}} compares the multicast rate performance of the proposed MA-aided system against FPA benchmarks across varying transmit power levels, $P$. As shown, the proposed algorithms consistently outperform FPA-based systems at identical power levels. Notably, this performance gap widens for smaller $N$, as the limited beamforming gain in such cases amplifies the benefits of MA position optimization. For larger $N$, beamforming gains dominate, reducing the relative impact of MA positioning. {\figurename} {\ref{fig_Rate_User}} plots the multicast rate versus the number of users. Both MA and FPA systems experience rate degradation as user count increases, owing to diminishing spatial degrees of freedom for multicast beamforming. However, the MA system maintains superior performance across all user densities, leveraging dynamic antenna positioning to better align with heterogeneous channel conditions. These results validate the efficacy of the proposed algorithms and underscore the inherent advantages of MAs over FPAs in multicasting.
\subsection{Two-User Case}

\begin{figure}[!t]
 \centering
\setlength{\abovecaptionskip}{0pt}
\includegraphics[width=0.35\textwidth]{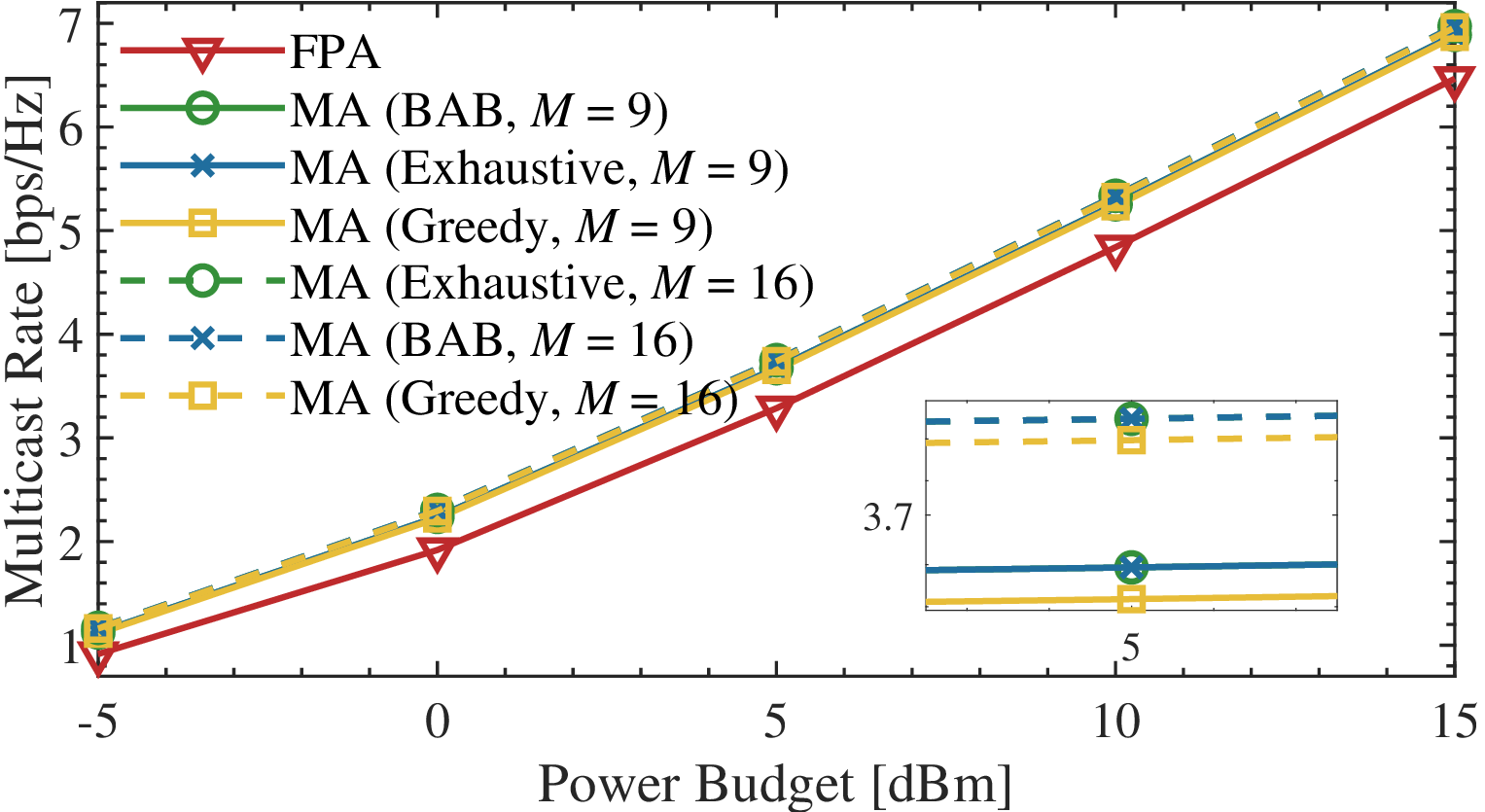}
\caption{Multicast rate vs. the power budget. $K=2$.}
\label{fig_MR_SNR}
\end{figure}

To validate the BAB-based optimal design, we analyze a two-user LoS scenario with $L_1=L_2=1$ and $\lvert\sigma_{1,1}\rvert^2=\lvert\sigma_{1,2}\rvert^2=\frac{\mu_1}{L_1}=\frac{\mu_2}{L_2}$. 

\begin{figure}[!t]
 \centering
\setlength{\abovecaptionskip}{0pt}
\includegraphics[width=0.35\textwidth]{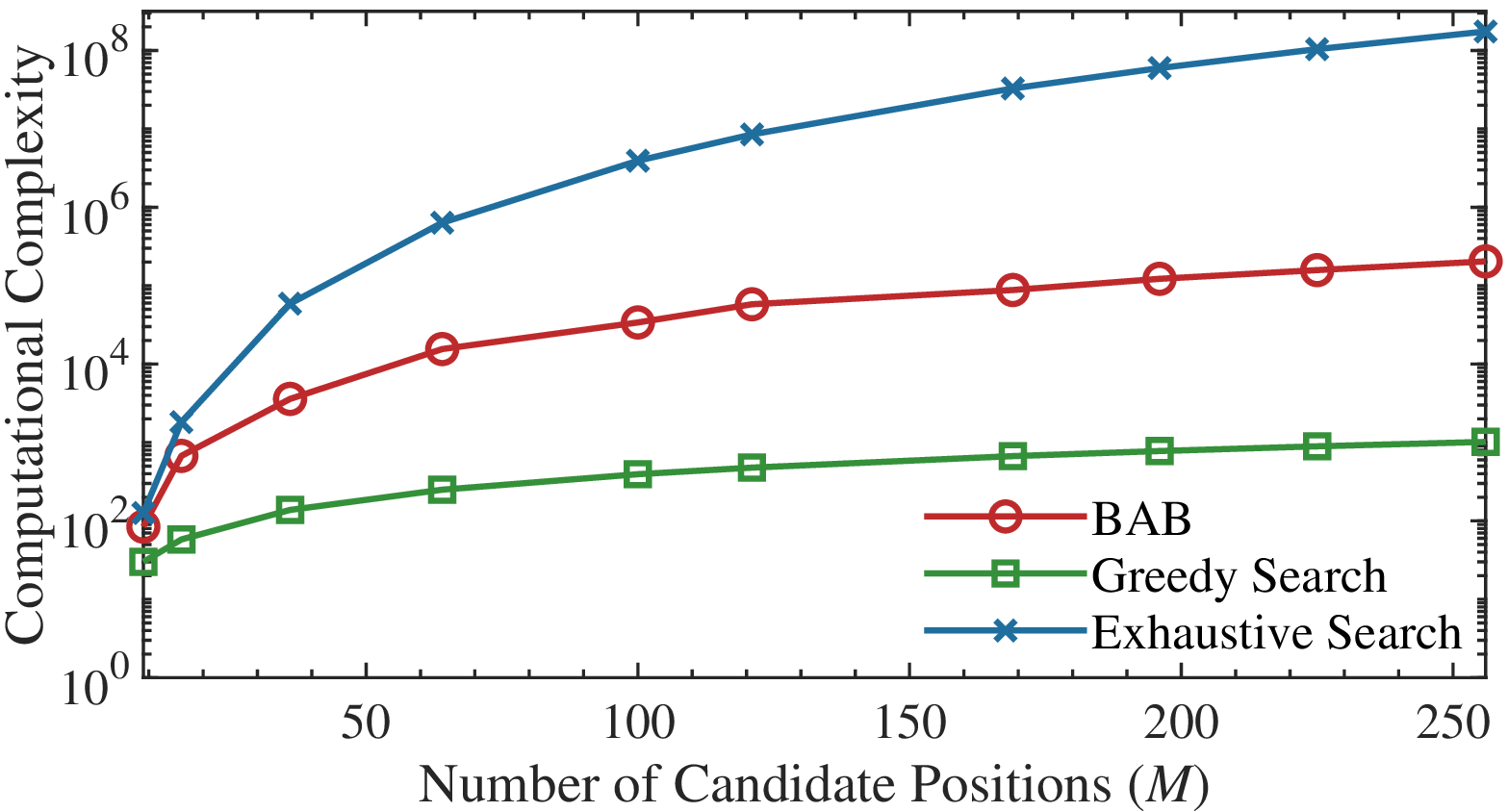}
\caption{Complexity vs. the number of candidate positions.}
\label{fig_Complexity}
\end{figure}

{\figurename} {\ref{fig_MR_SNR}} compares the multicast rates achieved by the BAB-based method, greedy search, and FPA benchmarks across varying transmit power $P$. The proposed BAB algorithm outperforms the FPA baseline at identical power levels, with the performance gap widening as the number of candidate positions $M$ increases. Notably, the advantages of MAs diminish slightly compared to the multipath model in Section \ref{Section:Numerical Results:General Case}. This occurs because LoS-dominated far-field channels restrict MA adjustments to phase shifts, which limits their ability to exploit multipath diversity. For reference, the multicast rate achieved by exhaustive search is also presented. We note that the exhaustive search achieves identical rates to the BAB method, while the greedy search exhibits negligible performance loss. 

{\figurename} {\ref{fig_Complexity}} evaluates computational complexity, measured by the number of visited nodes in the search tree, versus $M$. The BAB method reduces complexity significantly compared to exhaustive search but remains more computationally intensive than the greedy approach. These results demonstrate that the greedy search balances performance and complexity effectively, making it suitable for large-scale systems.
\section{Conclusion}
In this paper, we investigated the joint optimization of transmit beamforming and antenna positioning to maximize the multicast rate in MA-assisted MISO systems. For the general multiuser scenario, we developed an AO framework based on SCA. In the simplified two-user case, we introduced both a low-complexity greedy search algorithm and a BAB method, which can either closely approximate or achieve the optimal performance bound. Numerical results were presented to validate the effectiveness of the proposed algorithms, demonstrating that the use of movable antennas significantly enhances multicast performance compared to conventional FPA systems.

\end{document}